\documentclass{aa}
\usepackage{hyperref}
\bibpunct{(}{)}{;}{a}{}{,}
\usepackage{latexsym}
\usepackage{natbib}
\usepackage{graphicx}
\usepackage{times}  \DeclareSymbolFont{operators}{OT1}{ptm}{m}{n}
\usepackage{color}

\newcommand{\sect}[1]{Sect.\,\ref{#1}}
\newcommand{\sects}[1]{Sects.\,\ref{#1}}
\newcommand{\fig}[1]{Fig.\,\ref{#1}}

\newcommand{\tab}[1]{Table\,\ref{#1}}
\newcommand{\equ}[1]{Eq.\,(\ref{#1})}
\newcommand{\ang}[1]{#1\,\AA}
\newcommand{\abs}[1]{\lvert #1\rvert}

\graphicspath{{./}{figs/}}
\usepackage{amstext}

\begin{document}

%
\title{Using coronal seismology to estimate the magnetic field strength in a realistic coronal model} 

\titlerunning{Estimate the magnetic field strength in a realistic coronal model}
\authorrunning{F.~Chen \& H. Peter}

\author{F.~Chen and H.~Peter}
\institute{Max-Plank-Institut f\"ur Sonnensystemforschung, 
              37077 G\"ottingen, Germany
}

\date{}

\abstract%
%
{}
{Coronal seismology is extensively used  to estimate properties of the corona, e.g. the coronal magnetic field strength are derived from oscillations observed in coronal loops. We present a three-dimensional coronal simulation including a realistic energy balance in which we observe oscillations of a loop in synthesised coronal emission. We use these results to test the inversions based on coronal seismology.}
{From the simulation of the corona above an active region  we synthesise extreme ultraviolet (EUV) emission from the model corona. From this we derive maps of line intensity and Doppler shift providing synthetic data in the same format as obtained from observations.
We fit the (Doppler) oscillation of the loop in the same fashion as done for observations to derive the oscillation period and damping time. 

}
{The  loop oscillation seen in our model is similar to  imaging and spectroscopic observations of the Sun. The velocity disturbance of the kink oscillation  shows an oscillation period of 52.5\,s and a damping time of 125\,s, both being consistent with the ranges of periods and damping times found in observation. Using standard coronal seismology techniques, we find an average magnetic field strength of $B_{\rm kink}{=}79$\,G for our loop in the simulation, while in the loop the field strength drops from some 300\,G at the coronal base to 50\,G at the apex.
Using the data from our simulation we can infer what the average magnetic field derived from coronal seismology actually means. It is close to the  magnetic field strength in a constant cross-section flux tube that would give the same wave travel time through the loop.}
{Our model produced not only a realistic looking loop-dominated corona, but also provides realistic information on the oscillation properties that can be used to calibrate and better understand the result from coronal seismology.}

\maketitle

\section{Introduction}\label{S:intro}

Loops dominate the appearance of the corona of the Sun. In particular in active regions observed at extreme ultraviolet (EUV) wavelengths they are seen as fine threads outlining the magnetic field. In response to a significant localized energy deposition, like in a flare, these loops can bee seen to oscillate \citep{Nakariakov+al:1999}. Since then oscillations in the corona have been studied extensively in terms of theoretical investigations, numerical models, and observations \citep[e.g. reviews by][]{Nakariakov+Verwichte:2005,Banerjee+al:2007,Wang:2011,DeMoortel+Nakariakov:2012}. These oscillations of coronal loops are used for diagnostics of corona loops, in particular of the magnetic field strength. This was proposed already  by \citet{Uchida:1970} and \citet{Roberts+al:1984} who provided the theoretical ground work. Traditionally EUV imaging and spectroscopy gave access only to plasma properties, i.e., temperature, density, abundance and flows \citep[e.g.][]{Mariska:1992}. Coronal seismology holds the key to infer information also on the magnetic properties, in particular for the field strength and resistivity. The key techniques for coronal seismology  give access to the mode of the oscillation and the phase speed of the corresponding wave. 

The oscillations of coronal loops have been investigated mainly by two techniques. The first technique  measures the displacements and disturbances in EUV images \citep{Aschwanden+al:1999,Nakariakov+al:1999,DeMoortel+al:2000,Nakariakov+Ofman:2001,
Aschwanden+Schrijver:2011,Yuan+Nakariakov:2012,Verwichte+al:2013b,Guo+al:2015}. The second technique  investigates the periodic patterns in the Doppler shifts obtained from EUV spectrometers \citep{Ofman+Wang:2002,Wang+al:2003,Wang+al:2007,VanDoorsselaere+al:2008,
Erdelyi+Taroyan:2008,Ofman+Wang:2008,Mariska+Muglach:2010,Tian+al:2012}. Both techniques give an \emph{average} magnetic field strength of typically 10\,G to 100\,G in the loop. This  is in general  consistent with the field strength derived from extrapolations of the photospheric magnetic field \citep[e.g.][]{Schriver+al:2006,Wiegelmann+Sakurai:2012}. However, the results from coronal seismology can only provide some \emph{average} value of the magnetic field along the loop, while clearly the magnetic field has to expand with height and therefore the magnetic field strength will change along the loop. Consequently, it remains unclear what this average derived from coronal seismology really means. 

While the original work was assuming a rather simple setup \citep[e.g.][]{Roberts+al:1984}, more recent theoretical efforts have accounted also for the more complex structure of the real Sun, e.g., curved geometry, density stratification, or non-uniform cross section \citep{VanDoorsselaere+al:2004,Verwichte+al:2006,Erdelyi+Verth:2007,
Arregui+al:2007,Goossens+al:2009,Ruderman+Erdelyi:2009,Selwa+al:2011}. In a 3D model \citet{DeMoortel+Pascoe:2009} investigated the estimate of the magnetic field strength by coronal seismology. In their model the magnetic field strength and number density along the coronal loop are assumed to be constant \citep{Pascoe+al:2009}. This configuration sets a single reference value, and the authors did find a difference between their reference value and field strength derived from the oscillation of the model structure.

To make things worse, in the real corona the magnetic field strength varies along the loop, as does the number density, despite of the large density scale-height in the corona, and this might have a significant impact on the results of coronal seismology \citep{Ofman+al:2012}. \citet{Aschwanden+Schrijver:2011} and \citet{Verwichte+al:2013} compared the magnetic field strength from coronal seismology with those obtained from potential or force-free extrapolations of the magnetic field in the same structures. Because of the limitation of the approaches (i.e. potential or force-free), it is uncertain if the extrapolated magnetic field line actually match the observed loops. Promising examples were presented by \citet{Feng+al:2007} who showed that the magnetic field lines in a linear-force-free extrapolation do follow the loop structures reconstructed from stereoscopic observations.

Given all the difficulties of getting the reference values from observations, it is  highly desirable to test coronal seismology in a model corona, which has realistic plasma properties and a magnetic field configuration similar to that of a real active region. Forward coronal models \citep{Gudiksen+Nordlund:2005a,Gudiksen+Nordlund:2005b,Bingert+Peter:2011} account for the cooling through the optically thin radiation and the highly anisotropic heat conduction along magnetic field lines, and successfully reproduce a variety of coronal phenomena \citep{Peter:2015}. In those models the the horizontal motions in the  photosphere induce currents in the corona, either through fieldline braiding \citep{Parker:1972} or flux-tube tectonics \citep{Priest+al:2002}. The Ohmic dissipation of these currents  is sufficient to heat the coronal plasma to over one million K. The energy distribution in this type of model is consistent with the expectation of the nanoflare mechanism \citep{Bingert+Peter:2013}. The full treatment of the energy balance allows these models to resemble the plasma properties in the real corona, so that the synthesised emission from these models can be directly compared with real observations. These model successfully explain some basic features of coronal loops \citep[e.g. the non-expanding cross section,][]{Peter+Bingert:2012}. A one-to-one data-driven simulation can reproduce the appearance and dynamics in the particular solar active region that drives the simulation \citep{Bourdin+al:2013}.

Being successful in modelling the plasma properties and general dynamics of the corona, a further challenge is how well the loop oscillation in a realistic model resembles that in real observations. We analyse the coronal emission synthesised from a 3D MHD model in which we clearly see  loop oscillations. Analysing the synthetic data in the same way as observed coronal oscillations, we estimate the \emph{average} magnetic field strength in the loop. Having access to the full 3D cube of the simulation data, we can compare this \emph{average} value to the \emph{actual}  magnetic field strength that varies along the model loop. This comparison is a complementary to previous numerical experiments \citep{DeMoortel+Pascoe:2009} and observations \citep{Aschwanden+Schrijver:2011,Verwichte+al:2013}, and helps to better understand the implications  derived from the field strength inferred from  coronal seismology.

\section{Model setup}\label{S:model}

The coronal model we analyse here is based on the modelling strategy as described by \citet{Bingert+Peter:2011}.   The model setup is the same as in our previous model described in \citet{Chen+al:2014,Chen+al:2015.nat}. However, the simulation described here has significantly higher resolution: The $147\times74\times50$\,Mm$^3$ volume is now resolved by $1024\times512\times256$ grid points, which is an increase by a factor of 4 in each horizontal direction. The grid spacing is uniform in the horizontal direction (144\,km grid spacing) and non-uniform in the vertical direction (smoothly changing from 30\,km in the photosphere to 190\,km in the coronal part, ranging from 2\,Mm to 40\,Mm).
To solve the full MHD equations we use the Pencil Code \citep{Brandenburg+Dobler:2002}.\footnote{See also http://pencil-code.nordita.org.}

Our model corona is driven by the emerging magnetic field and the flows in the photosphere. These are taken from a simulation of the emergence of a magnetic flux tube from the upper convection zone through the photosphere \citep{Cheung+al:2010}. Here we use a model  where the emerging flux tube has no imposed twist \citep{Rempel+Cheung:2014}. In the process of the emergence  a pair of sunspots is formed by coalescence of small magnetic patches. This flux emergence model covers only a small part of the the photosphere. We take the time-dependent output of the flux emergence simulation at the solar surface (magnetic field, velocity, density and temperature) and impose this at the lower boundary of our coronal model \citep[just as in][but now at higher resolution]{Chen+al:2014}.

The major limitation for the time step in the explicit time-stepping scheme of our model is due to the Spitzer heat conduction. Therefore we evolve the equations in an operator-splitting manner. The governing equations without the heat conduction term are evolved by a regular 3rd-order Runge-Kutta scheme. We then evolve an  equation for the the heat conduction term alone in a sub-cycle. In the sub-cycle, we use a super-time-stepping scheme \citep{Meyer+al:2012}, which allows us to evolve the energy equation with a time step much larger than that for the original Runge-Kutta scheme.

The second-smallest time step in the simulation is (mostly) the Alfv\'en time step. The Alfv\'en speed in the model corona can be very large (up to 15\,000\,km/s), which would then limit the time step. Fortunately these large speeds are found only in a small fraction of the computational domain. To overcome this, we control the Alfv\'en speed $ v_{\rm A}$ by limiting the Lorentz force in the same way as \citet{Rempel+al:2009}. By multiplying the Lorentz force in the momentum equation by a correction factor $f_{\rm A}$ we ensure that the resulting effective Alfv\'en speed in the model, 
\begin{equation}\label{E:eff.Alfen.speed}
\tilde{v}_{\rm
A} = f_{\rm A}^{\,1/2} \, v_{\rm A},
\end{equation}
is always smaller than a maximum speed $v_{\rm{max}}{=}2000$\,km\,s$^{-1}$. As in \citet{Rempel+al:2009} the correction factor is defined as \begin{equation}
f_{\rm A}~=~v^2_{\rm{max}}\left(v^4_{\rm A}+v^4_{\rm{max}}\right)^{-1/2} .
\end{equation}
For a typical coronal loop with a number density of 10$^{9}$\,cm$^{-3}$ and a magnetic field strength of 50\,G at its  apex,  the  Alfv\'en speed (${\approx}3000$\,km\,s$^{-1}$) is reduced by about a factor of 1.5. Significant corrections will mostly occur in low-density regions (where the coronal Alfv\'en speed gets large), which will not appear bright in the synthetic images and thus can be expected to play a minor role only for the analysis we present in this paper. Despite
this limiting of the Alfv\'en speed, in the model corona the plasma-$\beta $ will still be well below 1 and the effective Alfv\'en speed will still be  an order of magnitde larger than the maximum sound speed (250\,km s$^{-1}$at 3\,MK).
In summary, this limiting of the Alfv\'en speed will have a minor effect only on our results, but it will provide a significant speed-up of the numerical simulation.

Together, the treatment of the heat conduction and the Alfv\'en speed, introduced only to speed-up the simulation time, provide a speed-up of typically a factor of five, at little extra computational cost.

Our high-resolution simulation is able to resolve  small-scale photospheric magnetic structures and flows related to the granular motions in the photosphere which are used as input from the flux-emergence simulation. The interaction of the plasma flow and magnetic field in the model photosphere, especially at the outer edge of strong flux concentrations, produces the enhanced upward Poynting flux that can bring enough energy into the upper solar atmosphere and power the coronal loops. This energy flux sustains a more than 3 MK hot corona at pressures according to the classical scaling laws \citep{Rosner+al:1978}.

\begin{figure}
\includegraphics{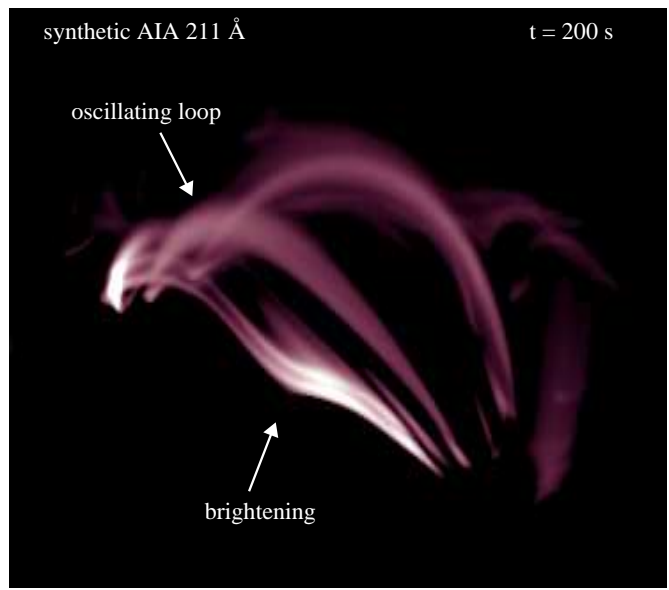}
\caption{System of coronal loops synthesised from 3D MHD model. This displays  a snapshot of the model corona as it would appear in an EUV image taken by AIA in the \ang{211} channel (in logarithmic scale). It is dominated by emission from \ion{Fe}{XIII} showing  plasma at around 2\,MK. The distance between the two footpoints of the loop system is about 35\,Mm and the loops have lengths of about 45\,Mm to 50\,Mm. The arrows indicate the position of the brightening that triggers the oscillation (see \sect{S:trigger}) and the oscillating loop (\sect{S:measure}). The full temporal evolution over 54\,minutes is available as a movie in the online edition.
The movie starts early in the simulation, when there is no coronal emission. In response to the flux emergence coronal loops form, and at about 72.5\,minutes a trigger sets the oscillation in motion. Then a second counter in the movie shows the time in s starting at 72.5 minutes,  consistent with the time used in \fig{F:damp_osci}.%
\newline
The movie is also available at\newline
\href{http://www2.mps.mpg.de/data/outgoing/peter/papers/2015-osci/movie-fig1.mp4}%
     {http://www2.mps.mpg.de/data/outgoing/peter/papers/2015-osci/movie-fig1.mp4}.
\label{F:rotate}}
\end{figure}

We show in \fig{F:rotate} a snapshot of a time series of the synthesised EUV images according to the response function of the \ang{211} channel of the Atmospheric Imaging Assembly \citep[AIA,][]{Boerner+al:2012}. Here we follow the procedure outlined in \citep{Peter+Bingert:2012} to calculate the AIA emission  expected from the model corona. In these synthetic images we find numerous bright EUV loops, which represent the coronal plasma at temperatures around  2\,MK. In these loops the number density is about 10$^9$\,cm$^{-3}$. The properties of the coronal plasma are in quantitative agreement with our previous simulation, and they are also consistent with typical values derived from real observations.

Beyond the appearance in a synthetic EUV snapshot, the model also captures the dynamic nature of the real corona to some extent. The animation associated with \fig{F:rotate} shows that bright features can show up or disappear within a few minutes in a certain EUV passband. This is consistent with modern observations, for example from AIA.

\section{Oscillation observed in synthesised coronal emission}\label{S:obs}

\subsection{Trigger of the oscillation}\label{S:trigger}

Similar as in our previous model of an emerging active region \citep{Chen+al:2014,Chen+al:2015.nat} loops form in the active region in response to  the heating driven by  footpoint motions in the photosphere.  This leads to a more or less continuous evolution of the  loops: some loops form, others fade away. A snapshot of the loops in coronal emission is displayed in \fig{F:rotate}, the temporal evolution is shown in the attached movie.
In the following we relate all times relative to the time 72.5\,minutes after the actual start of the simulation.

At the time   $t{\approx}$200\,s some loops start to oscillate just after a strong transient brightening of some close-by loops. These locations are indicated in the snapshot in  \fig{F:rotate} as well as in the attached movie. 

We  traced the cause of the strong transient brightening as being due to an enhancement of  Ohmic heating along the field lines at the flank of the active region. The cool plasma is quickly heated to a temperature at which the response function of the AIA \ang{211} channel peaks. The increase of the  heating is caused by an increase of the currents in response to  reconfiguration of the coronal magnetic field driven by photospheric flows in the periphery of the sunspots (as the result of the flux emergence). The strong brightening associated with this increased heating rate can be considered as a very small flare on the Sun (even though we do not imply here that this is a flare model).

The increased currents come along with a transient increase of the Lorentz force, leading to a kick  perpendicular to the field lines. The disturbance propagates from the brightening site across the active region and triggers the transverse oscillation of the nearby loops. Again, this can be considered in analogy to eruptions on the real Sun, which are suggested to be the main trigger mechanism of observed loop oscillations. This oscillation is clear, albeit not violent, in the   movie attached to  \fig{F:rotate}.

The travel time of the triggering disturbance from the transient brightening to the farthest visible loop is about 12\,s (propagation with an Alfv\'en speed of almost 2000\,km\,s$^{-1}$ across a distance of some 25\,Mm). This is why at the cadence of 10\,s for the movie shown with \fig{F:rotate} the oscillation of the loop seems to start at almost the same time as the transient brightening triggering the event.

The actual process triggering the oscillation is a very interesting process in itself, but it is not the main interest of this study. In contrast, we concentrate here on the observable consequences of this event and investigate to what extent the methods of coronal seismology can recover the atmospheric conditions at the location of the oscillating loop, in particular the magnetic field.

\subsection{Imaging and spectroscopy of model data}\label{S:spectral} 

The synthesised imaging data (movie with \fig{F:rotate}) reveal a transverse oscillation the loop that lasts
for about 300\,s and decays gradually. Its appearance is  similar to the oscillations widely found in EUV observations. Mostly this type of oscillation is
interpreted as the standing, fast kink mode \citep[e.g.][]{Nakariakov+Verwichte:2005}. Although the loop oscillation is  clear in the animation for the human observer, a quantitative analysis of this oscillatory displacement of the loop position is not so clear, because the amplitude of the oscillation is only of the order of or even smaller  than the width of the coronal loop.

Alternatively, one can investigate the velocity disturbances through Doppler shifts from (synthesised) spectroscopic observations. Because the loops are inclined, when observed from straight above the  transversely oscillating loops will result in periodic Doppler shifts. To study this, we simulate an observation from the top of the active region, i.e. as if we would observe an active region at the disk center. For this we calculate the emission line profile at each grid point based on the output of the MHD simulation and then integrate along a vertical line-of-sight. This procedure follows  \citet{Peter+al:2004,Peter+al:2006}. To be consistent with the imaging data shown in \fig{F:rotate}, we study line profiles from the same ion and choose the   \ion{Fe}{XIII} \ang{line at 202} line that has been observed abundantly with the EUV Imaging Spectrometer \citep[EIS;][]{Culhane+al:2007}.

\begin{figure}
\includegraphics{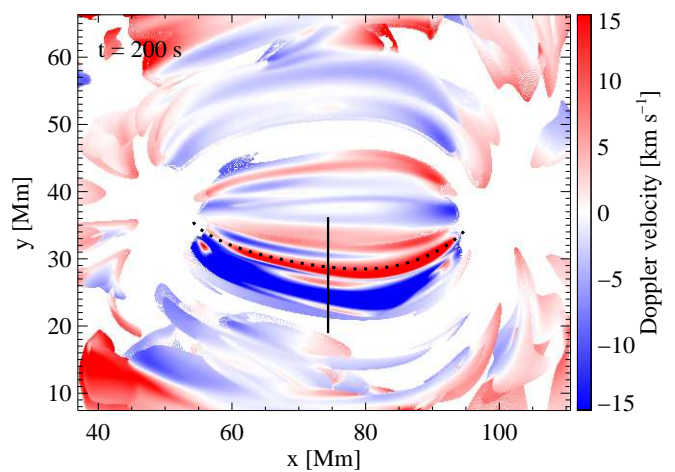}
\caption{Map of Doppler shifts of synthesised \ion{Fe}{XIII} \ang{202} data. This shows the active region seen from straight above, i.e. the line-of-sight is vertical, corresponding to an observation at disk center. The black solid line indicates the position of the slit used to simulate a sit-and-stare observation for the analysis of the Doppler oscillation  (see \fig{F:damp_osci} and \sect{S:spectral}). The dotted line indicates the location of the oscillating loop as seen in the movie attached to \fig{F:rotate}.
\label{F:dop_vel}}
\end{figure}

The   \ion{Fe}{XIII} profiles are mostly close to a single Gaussian. Their width is determined by the local plasma temperature and the distribution of the line-of-sight velocities. To derive the line intensity and shift we take the zeroth and first moment of the line profile. The resulting Doppler  map in \fig{F:dop_vel} basically shows the vertical (line-of-sight) velocity of the 2\,MK hot plasma in the active region (at the same time as the snapshot in \fig{F:rotate}). This Doppler map shows the line shifts at the same time across the whole map. In a real observation with a slit spectrometer one would have to produce a raster scan to obtain this map, e.g. EIS would need up to one hour to raster a field-of-view as shown in \fig{F:dop_vel} (depending on observation parameters). Thus a real observation would look quite different than this instantaneous Doppler map.

In a real observation usually one would perform a so-called sit-and-stare observation to catch the oscillation,
i.e., one would keep the slit at a fixed position in the active region and study the temporal evolution at that location. For our synthetic observations, we choose a  slit oriented along the $y$-direction located just in the middle between the footpoints of the loop (solid line in \fig{F:dop_vel}). The dotted line indicates the location of the loop that is seen to be oscillating in the intensity data (cf. movie with \fig{F:rotate}). Here we catch the loop in a phase of the oscillation moving away from the virtual observer.   The slit roughly crosses the apex of the loops.

For the further analysis we extract the line intensity and the Doppler shift along this slit from the synthesised spectral data with a cadence of 10\,s. This roughly matches the typical cadence used in real observations. The resulting time series of this synthetic observation is  comparable to real data acquired from active region loops in a sit-and-stare observation.

\subsection{Quantitative analysis of the loop oscillation}\label{S:measure}

\begin{figure}
\includegraphics{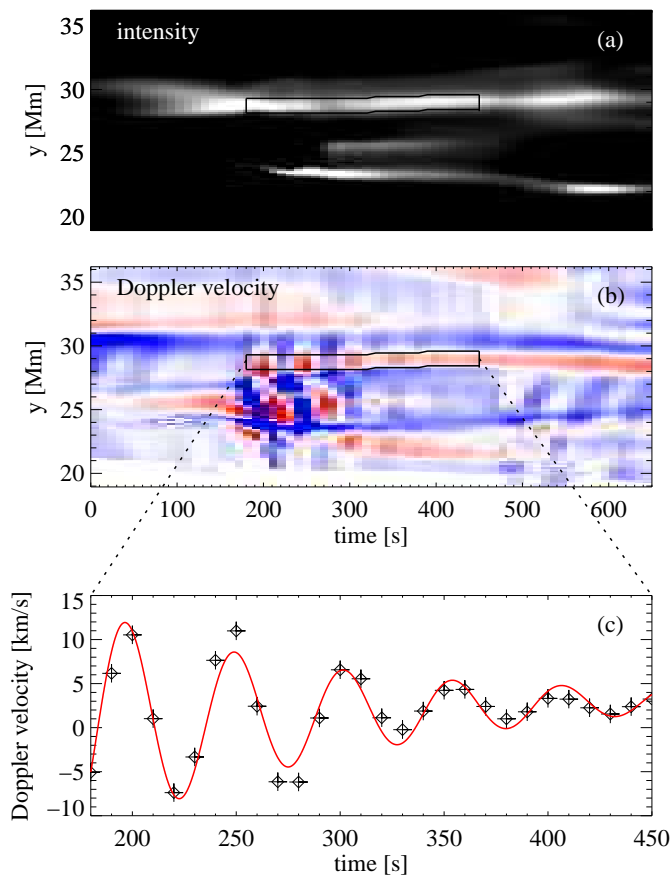}
\caption{Oscillation of the loop. Panels (a) and (b) show the intensity and Doppler shift of the \ion{Fe}{XIII} line synthesised from the coronal model as a function of time and space along the slit. The location of the slit is indicated in \fig{F:dop_vel} and the range of Doppler shifts is from $-15$\,km\,s$^{-1}$ to $+15$\,km\,s$^{-1}$, just as in \fig{F:dop_vel}. The black boxes in these space-time plots indicate  the position of the loop with the clearest oscillation pattern. Panel (c)  shows    the Doppler shift of the loop as a function of time. Here the Doppler shifts are averaged along the slit across the loop, i.e. along $y$ within the black box in panel (b). The red line shows the fit by a damped sinusoidal function as defined in  \equ{E:osci}. See \sect{S:measure}.
\label{F:damp_osci}}
\end{figure}

To analyse the oscillation we first check time-space diagrams for the line intensity and Doppler shift. Just like for sit-and-stare observations of the real Sun \citep[e.g.][]{Wang+al:2009,Mariska+Muglach:2010} these show the line intensity and shift  as a function of time and space (along the slit). For the slit position indicated in  \fig{F:dop_vel} these are displayed in panels (a) and (b) of \fig{F:damp_osci}.

The line intensity does not show a clear oscillation in the $y$-direction in this diagram. This is consistent with the impression from the synthetic AIA images that the amplitude of the oscillation (in space) is smaller than the loop width. In addition,  the loop is inclined and the oscillation is transverse, which further reduces the amplitude of loop displacement.

 In contrast, the  oscillation is very clear in Doppler shift (\fig{F:damp_osci}b). It starts at around $t{=}$200\,s, which is consistent with the oscillation seen as a slight displacement of the loop in coronal emission (see movie with \fig{F:rotate}). The oscillatory pattern is seen over the whole field-of-view shown in \fig{F:damp_osci}b, underlining the impression from the intensity images that the trigger leads to a disturbance of a whole arcade of loops. While three distinct loops can be identified in \fig{F:damp_osci}\,(a) near $y{=}$23\arcsec, 26\arcsec, and 29\arcsec,
in the following we will concentrate on the latter one. This shows the clearest oscillating pattern that lasts from ca.\ $t{=}$180\,s to 450\,s and is marked by a black box in \fig{F:damp_osci} (a)
and (b).

To get a clear signal, we average the Doppler shift in the oscillating loop in the black box in \fig{F:damp_osci} (b) in the $y$-direction. The resulting (mean) Doppler velocity as a function of time is shown by the black symbols in \fig{F:damp_osci} (c) which clearly shows a  damped oscillation. To extract the period $P$ of the oscillation and its damping time $\tau$ we fit this oscillation by an exponentially damped sinusoidal function,
\begin{equation}\label{E:osci}
f(t)~~=~~A_0~\exp {\bigg[}{-}\frac{t-t_0}{\tau}{\bigg]} ~ \sin{\bigg[}\frac{2\pi}{P}(t-t_0){\bigg]} ~~+~~ A_1 t ~~+~~A_2~.
\end{equation}
Here $t$ is the time, $t_0$ the initial time, and $A_0$ the amplitude of the exponentially damped sinusoidal function. In addition, $A_1$ and $A_2$ account for a linear background. Fitting the Doppler shifts in \fig{F:damp_osci} (c) with a uniform weight we obtain a period and a damping time of
\begin{equation}\label{E:times}
\begin{array}{rcr@{}l}
P    & = & 52&.5\,\mbox{s}\\
\tau & = & 125&\,\mbox{s}
\end{array}
\end{equation}
We applied the same analysis also for  some other slit positions. When the slit is  located  midway from the loop footpoint to the apex, the average Doppler perturbation in the loop appears very similar to that at the apex, albeit with a reduced amplitude. Still, the damping time and period derived at these other positions  are fully consistent with the results at the loop apex.
If the slit is positioned close to the loop footpoints, the oscillatory Doppler signal gets mixed with  slowly changing Doppler shifts due to  flows (along the magnetic field) filling and draining the loop \citep[cf.\ ][]{Chen+al:2014}.  Therefore, when analysing the oscillation close to the footpoints we first remove this  smoothly varying signal before fitting the oscillation by \equ{E:osci}. Again the period and damping time are consistent with the values found near the apex listed in \equ{E:times}. This underlines that the loop shows a coherent oscillation all along (see also \sect{S:mode}).

Comparing these values of \equ{E:times} to observations, they are found at the lower end of
the distribution found in the compilation of oscillations by \citet{Verwichte+al:2013b}.
This is not surprising, because the the size of our computational box is
limited, and thus the loops we can study have lengths of below 50\,Mm. The
observed loops are mostly longer by a factor of two (or more). Hence the
expected period would be longer by that factor (the period is proportional
to the loop length, see \sect{S:mag_inv}.

An important observational test for our model loop is the scaling between
period and damping time as derived by \citet{Verwichte+al:2013b}
based on a sample of 52 oscillating loops. They found  $\tau=\alpha
P^{\gamma}$, with ${\rm log_{10}}\,\alpha = 0.44{\pm}0.31$ and $\gamma=0.94{\pm}0.12$.
The period  and damping time
of the oscillation found in  our model listed in \equ{E:times} fits this scaling relation very well.
This suggests that our model for the loop oscillation and its damping capture
the right physical processes, even though the period and damping time are
at the lower end of what is found in observations.

\subsection{Mode of the oscillation}\label{S:mode}

Apart from the measurement of the oscillation parameters, the identification
of the corresponding wave mode is equally important. In the synthetic EUV
images the loop oscillates primarily in the transverse direction without
any visible (stationary) node along the loop (cf.\ movie
with \fig{F:rotate}). This is supported by the Doppler patterns that have
the same sign all along the loop (see \fig{F:dop_vel}). 
Together this suggests that the oscillation in the model is a fundamental kink mode.

A quantitative test for the presence of the fundamental kink mode is given through the phase difference of the velocity disturbance measured at different positions along the loop (D. Yuan, private communication). Because in the fundamental kink mode the loop should oscillate coherently all along, this phase difference should vanish.  

\begin{figure}
\includegraphics{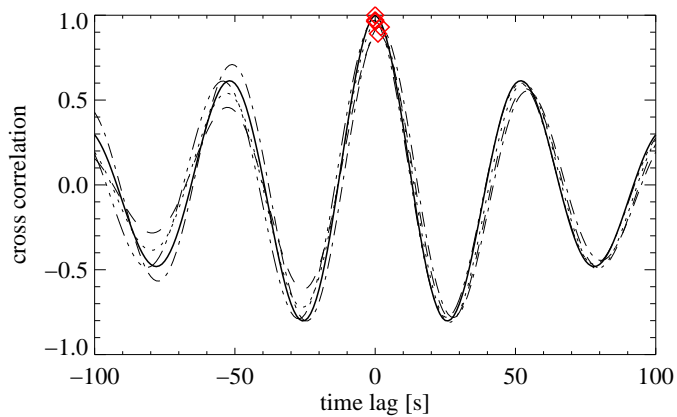}
\caption{Cross correlations of the oscillation at different locations along the loop. The five curves show the cross correlation of the loop apex with five positions along the loop (apex, two positions midway from apex to loop, and two near the footpoint at both sides). The solid line shows the self-correlation at the loop apex. The range of the time lag shown here is  about four times the oscillation period of 52.5\,s at the loop apex. The red diamonds indicate the peaks of the cross correlation functions. See \sect{S:mode}.
\label{F:cross}}
\end{figure}

To estimate the phase difference between the velocity disturbance at different positions along the loop we cross correlate the oscillation at the apex to four other locations, two midway to the footpoints and two close to the footpoints at each side. For this we use the fits to the variation of the Doppler shifts from \equ{E:osci} and show the cross correlation as a function of the time lag in \fig{F:cross}.  All cross correlations peak at about zero time lag, i.e.\ the velocity disturbances of the oscillation at different positions have no phase difference \footnote{For kink mode oscillations, there is a phase difference of $\pi/2$ between the velocity disturbance and the displacement at the same position of the loop.}. Moreover, the peak values of the cross correlations are close to unity, indicating that the oscillations are very similar at different positions, i.e. that the loop oscillates as a whole.

\begin{figure}
\includegraphics{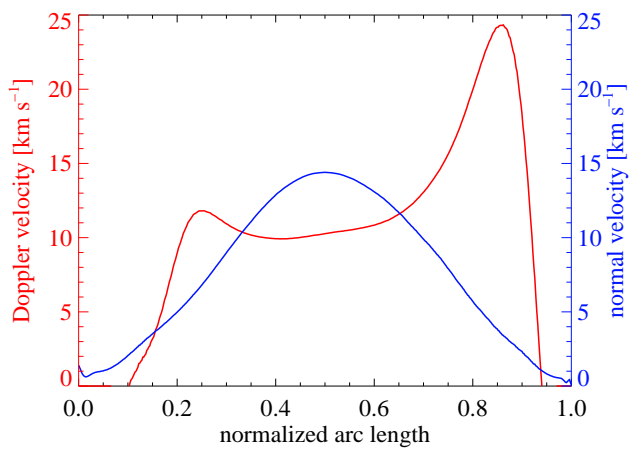}
\caption{Line-of-sight Doppler shift and normal velocity of the oscillation along the loop. The blue curve shows the velocity of the oscillating loop in the normal direction at a snapshot of maximum velocity amplitude (at $t{=}200$\,s). The red line displays the Doppler shifts from the synthesised spectral profiles, basically the vertical velocity in the loop, which also show signs of the field-aligned flows near the footpoints. The arc length has been normalized to the loop length, i.e. the footpoints in the photosphere are at 0 and 1, the apex is around 0.5. See \sect{S:mode}.
\label{F:vvlos_line}}
\end{figure}

Another indicator for the fundamental mode is given thorough the variation of the normal velocity along the loop. Because for the fundamental mode two nodes of the oscillation should be found at the footpoints, there should be a smooth drop of the normal velocity from the apex to the feet. This is confirmed by the plot in \fig{F:vvlos_line} which shows the normal velocity at a time near maximum velocity. There we also illustrate the difficulties that would be encountered in real solar observations for this test. While in our model we have direct access to the normal velocity, in observations this is not the case. To show this, in \fig{F:vvlos_line}
we also plot the Doppler shift as it changes along the loop when observing from straight above (i.e. the shift along the dotted line in \fig{F:dop_vel}). Essentially this shows the vertical velocity in the loop. While near the apex this basically reflects the line-of-sight component of the normal velocity, near the footpoints this signal is heavily contaminated by the flows in or out of the loop. Here we see a draining of the loop, i.e. downflows at both footpoints. The downflows (12\,km\,s$^{-1}$ and 24\,km\,s$^{-1}$) are comparable with the transverse oscillation speed (14 km s$^{-1}$) at the loop apex. In a real observations this might prevent any solid conclusions on the wave mode by checking the Doppler shift variation along the loop.

This result illustrates also an important constraint on simplified models for loop oscillations. Because one will never prevent such in- or outflows neat the loop footpoints in a complex enough model (we never see static loops in the 3D models) or on the real Sun, these flow structures have to be taken into account. In particular, because these flow speeds reach a good fraction of the sound speed.

Together with the the fact that the same periods and damping times are found all along the loop (\sect{S:measure}), the zero time-lag of the oscillation at different places along the loop and the clear existence of  nodes at the footpoints underlines that the oscillation in our model loop is indeed a fundamental kink mode.
\section{Oscillation frequency and magnetic field}\label{S:osci.frequency}

\subsection{Estimate of the magnetic field strength from coronal seismology}\label{S:mag_inv}

In a seminal paper \citet{Edwin+Roberts:1983} showed that the period of a loop oscillating in the corona would depend on the magnetic field, which in turn allows to estimate the field strength if one can identify the oscillation and determine its frequency. In an observation the phase speed $c_{\rm k}$ of a kink mode oscillation usually is estimated  by
\begin{equation}\label{E:ck_obs}
c_{\rm k}~=~\frac{2L}{P},
\end{equation}
where $L$ is the loop length and $P$ the oscillation period \citep[e.g][]{Nakariakov+al:1999}.
The loop length is determined from EUV imaging observations and the period by fitting the oscillation in the displacement of the loop position of the Doppler shift (e.g. as described in \sect{S:measure}). \tab{T:param} summarises these  parameters derived from our synthesised observation of the oscillating loop and other loop parameters and results.
The value we derive here for $c_k$ of 1730\,km\,s$^{-1}$ is still below  the limiting Alfv\'en speed for the simulation, see discussion with \equ{E:eff.Alfen.speed}. When we derive the magnetic field strength, we still consider the limiting in the analysis below in \equ{E:ck2}.

\begin{table}
\caption[]{Parameters of the oscillating loop.\label{T:param}}
\begin{tabular}{lccc}
\hline
\hline
\noalign{\smallskip}
parameter &  symbol & value & from \\
\noalign{\smallskip}
\hline
\noalign{\smallskip}
coronal loop length & $L$ & 45\,Mm & EUV \\
temperature & $T$ & 1.5\,MK & model \\
internal density & $n_{\rm i}$ & $5.7{\times}10^8$\,cm$^{-3}$ & model \\
external density ratio & $n_{\rm e}/n_{\rm i}$ & 0.12 & model \\
oscillation period & $P$  & 52.5\,s & \equ{E:osci}  \\
damping time & $\tau$ & 125\,s & \equ{E:osci} \\
kink mode speed & $c_{\rm k}$ & 1730\,km/s & \equ{E:ck_obs} \\
inverted $\abs{B}$ & $B_{\rm kink}$ & 79\,G & \equ{E:ck2} \\
average $\abs{B}$ & $\left\langle B \right\rangle$ & 92\,G & \equ{E:B_ave} \\
\noalign{\smallskip}
\hline
\hline
\end{tabular}
\tablefoot{The loop length is derived from the images of the synthesised
corona (EUV). The temperature, the density in the loop, and the ratio of the
density outside to the density inside the loop are derived from the 3D MHD
simulation (model). The other quantities are derived from the equations referred to
in the table.}
\end{table}

As derived by \citet{Edwin+Roberts:1983}, the phase speed of this kink mode in a slender magnetic flux tube with uniform density depends on the Alfv\'en speed
$v_{\rm{A}}$ and the density $\rho$ inside and outside the flux tube, 
\begin{equation}\label{E:ck1}
c_{\rm k}~=~\left(\frac{\rho_{\rm i}v_{\rm Ai}^2+\rho_{\rm e}v_{\rm Ae}^2}{\rho_{\rm i}+\rho_{\rm e}}\right)^{1/2}.
\end{equation}
Here the subscripts i and e refer to internal and external, i.e. to inside and outside the flux tube.

Because the corona is a low-$\beta$ plasma, it is usually assumed that the magnetic field strength $B_{\rm{kink}}$ (derived from the kink mode oscillation) inside and outside the loop is the same. In our MHD simulation we can confirm that this is indeed the case. Using the effective Alfv\'en speed in our model from \equ{E:eff.Alfen.speed} to account for the Alfv\'en speed limiting,  \equ{E:ck1} reads  
\begin{equation}\label{E:ck2}
c_{\rm k}~~=~~\frac{B_{\rm{kink}}}{~\sqrt{\mu_0\rho_{\rm i}}~}~~\left(\frac{2}{1+\rho_{\rm e}/\rho_{\rm i}}\right)^{\!\!1/2}~~\left[\frac{f_{\rm Ai}+f_{\rm Ae}}{2}\right]^{\!1/2}.
\end{equation}
The last term $[\cdots]$ is not present when observers analyse their data. It is an artifact of the Alfv\'en speed limiting we applied in our model (see \sect{S:model}) and consequently we have to correct for this here, too.

If observations provided the kink mode wave speed through \equ{E:ck_obs} and if the internal and external densities can be estimated from observations, too, then \equ{E:ck2} provides an estimate for the magnetic field strength in the loop based on the analysis of the kink mode, $B_{\rm{kink}}$. In absence of a reliable tool for density diagnostics when using imaging observations,  one often simply assumes a typical value for the coronal density, e.g. $10^9$\,cm$^{-3}$, and a typical density contrast of $\rho_{\rm{e}}/\rho_{\rm{i}}\,{=}\,0.1$.

\begin{figure}
\includegraphics{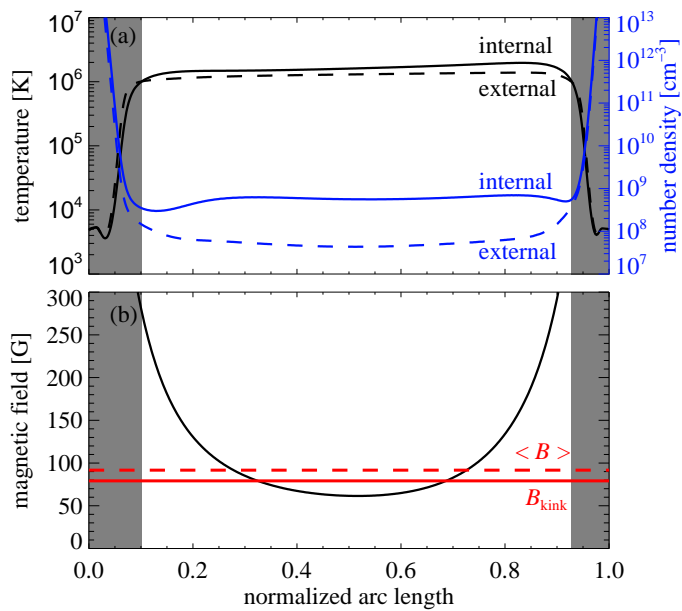}
\caption{Plasma parameters and magnetic field strength along the loop. Panel (a) shows the temperature (black) and density (blue) for the condition inside the EUV loop (solid lines labeled internal).  The corresponding values just outside the loop are plotted as dashed lines (labeled external). Panel (b) shows the magnetic field strength along the loop in the 3D MHD model. The red solid line shows the   coronal magnetic field strength derived from the kink oscillation, $B_{\rm kink}$, and the red dashed line the average magnetic field strength, ${\langle}B{\rangle}$, defined by \equ{E:B_ave}.  The position along the loop is normalized by the total loop length (53\,Mm). The positions 0.0 and 1.0 are at footpoints in the photosphere. The grey areas indicate the lower atmosphere, where $T{<}1$\,MK. The loop length in the coronal part, i.e. between the gray regions, is about 45\,Mm. See \sect{S:measure} for panel (a) and \sects{S:mag_inv} and \ref{S:compare} for panel (b).
\label{F:invert_B}}
\end{figure}

In our study we use the values for the internal and external densities as derived directly from the MHD model (see \tab{T:param}). For this we choose a group of magnetic field lines within the cross section of the oscillating loop as seen in EUV emission for the internal properties and another group of magnetic field lines in the ambient corona outside for the external properties. In \fig{F:invert_B} (a) we plot the number density and temperature averaged for the respective group of fieldlines along the loop, separately for the conditions inside and outside the loop.

In the coronal part (i.e.\ where $T{>}10^6$K) of the EUV loop, as well of the flux tube just outside the loop, the temperature and the density are roughly constant.
This is  because of the highly efficient heat conduction and the large pressure scale height at  high temperatures. In this coronal part (between the gray hatched areas of \fig{F:invert_B}), the average density inside the EUV loop is about $\rho_{\rm i}{=}3.8\times10^{-12}$\,kg m$^{-3}$ ($n_{\rm i}{=}5.7\times10^8$\,cm$^{-3}$). Outside the loop, the external average density is $\rho_{\rm e}{=}4.4\times10^{-13}$\,kg m$^{-3}$ ($n_{\rm e}{=}6.5\times10^7$\,cm$^{-3}$). The resulting density ratio is about $\rho_{\rm e}/\rho_{\rm i}\,{=}\,0.12$. The density we find in the loop is comparable to typically assumed values, and also the  ratio is  similar to the typical assumptions applied for tube models and for the inversion of observations \citep[e.g. $\rho_{\rm e}/\rho_{\rm i}\,{=}\,0.1$ by][]{Nakariakov+Ofman:2001}. 

The loop length in the coronal part is roughly 45\,Mm, which is consistent with the length one would derive from the synthesised EUV images. Together with the oscillation period of $P{\,=}\,52.5$\,s derived from fitting the oscillation in \sect{S:measure}, we obtain a kink-mode phase speed of $c_{\rm k}\,{=}\,1730$\,km\,s$^{-1}$ from \equ{E:ck_obs}. With the densities derived above we can use this value of $c_{\rm{k}}$ to derive the coronal magnetic field strength based on the kink mode oscillation from \equ{E:ck2} which yields $B_{\rm kink}\,{=}\,79$\,G.

This derivation of the average coronal magnetic field strength basically copies the procedure applied to observations. Now the main question is what this average actually means.

\subsection{Comparison with the actual magnetic field strength along the loop}\label{S:compare}

The magnetic field strength varies along the coronal loop in both the real corona and our numerical model, basically reflecting the expansion of the magnetic field. Therefore, the value deduced from coronal seismology does not necessarily represent the field strength at any particular position of the loop, but is an average through the coronal part of the loop. Most importantly, the significant variation of the magnetic field along the loop is usually neglected in coronal seismology, even though the importance of this effect has been shown before \citep[e.g.][]{Ofman+al:2012}. The profile of the magnetic field strength along the loop in our model (along the center field line of the loop) is plotted in \fig{F:invert_B} (b). In the the following we will investigate the field strength (e.g. its average) only in the coronal part, i.e. where $T{>}10^6$\,K between the gray areas in  \fig{F:invert_B}.

The arithmetic mean of the magnetic field strength along the loop is 120\,G, i.e.\ significantly higher than the value of $B_{\rm kink}\,{=}\,79$\,G derived from the oscillation. 
The reason for this discrepancy is that coronal seismology estimates the magnetic field strength from the observed average wave speed as defined as \equ{E:ck_obs}. Because the actual wave speed is (usually) not constant through the loop, this gives more weight to regions of low Alfv\'en speeds, i.e. to regions where the propagating wave packet dwells the longest. These are the regions of low magnetic field strength, which is why the  magnetic field strength derived from oscillations is much lower than an arithmetic mean.

To define an average that is closer to the value derived from oscillations, one has to account for the variation of the wave speed (or in other words, the wave travel time).
Similar to
\citet{Aschwanden+Schrijver:2011} we define
\begin{equation}\label{E:B_ave}
\left\langle B \right\rangle~=~L\,\left[\int \frac{ds}{B(s)}\right]^{-1},
\end{equation}
where $s$ is the coordinate along the arc length of the loop. \cite{Aschwanden+Schrijver:2011} argue that a flux tube with a field strength $B(s)$ changing along the loop would oscillate with the same frequency as a flux tube with a constant field strength $\langle B \rangle$  as derived from \equ{E:B_ave}. In our model loop, we find $\langle B \rangle{=}92$\,G, as indicated by the red dashed line in \fig{F:invert_B} (b). Compared with the arithmetic average along the loop, this is closer to the the magnetic field derived from the kink mode, $B_{\rm{kink}}{=79}$\,G,  but it is still significantly larger by some 15\% to 20\%.

That the average $\langle B \rangle$ based on \equ{E:B_ave} comes reasonably close to the value derived from the oscillation, $B_{\rm{kink}}$, underlines  that the magnetic field
strength estimated by coronal seismology represents the average magnetic field
weighted with the wave speed along the loop. Basically coronal seismology returns the value of the magnetic field that one would find in a flux tube with constant magnetic field with the same Alfv\'en crossing time.

However, the  field derived by coronal seismology, $B_{\rm kink}$, is still 15\% to 20\% smaller than the average $\left\langle B \right\rangle$, i.e.\ the true magnetic field in our simulation (or in the real Sun). This implies that the actual wave speed (in the model) is underestimated by the theoretical value from coronal seismology. A difference of up to 50\% between the derived magnetic field strength and the actual field strength was also found by \cite{DeMoortel+Pascoe:2009} who investigated a MHD model (albeit with a less realistic magnetic setup and much simpler thermodynamics). They concluded that this is because ``the combined effect of the loop curvature, the density ratio, and aspect ratio of the loop appears to be more important than previously expected'' \citep{DeMoortel+Pascoe:2009}.

Further more the (magnetic) complexity of coronal loops in our model and on the real Sun could lead to such a deviation. One possible explanation might be the aspect ratio (i.e. width/length) of the loop. The particular oscillating loop highlighted in \fig{F:rotate} has an aspect ratio of about 0.04 (2\,Mm width, 45\,Mm long; other loops in the model are thinner). In observations the aspect ratio is a bit smaller, typically 0.02 (2\,Mm width at 100\,Mm length). However, it has to be noted that here the width refers to the width of the EUV loop, while 3D models show that the corresponding density structures these EUV loops are embedded in are thicker \citep{Peter+Bingert:2012,Chen+al:2014}. Thus we can expect the aspect ratio to be several percent.

\citet{Edwin+Roberts:1983} noted that the actual kink mode speed is equal to $c_{\rm k}$ defined by \equ{E:ck1} only if the width of the flux tube is much smaller than the wavelength of the disturbance. Otherwise, it decreases with an increasing aspect ratio \citep[see Eq.\,15 in ][]{Edwin+Roberts:1983}. Therefore it might be that in our model (and on the real Sun) this assumption of  \citet{Edwin+Roberts:1983} is not fully applicable. Other effects, e.g. that plasma-$\beta$ is non-zero \citep[see Eq.\,11 for large $\beta$ of][]{Edwin+Roberts:1983}, a smooth density profile across the loop, or flows along the loop might play a role, too, in our model as well as on the real Sun. A detailed theoretical investigation of these effects is beyond the scope of this paper, but should be addressed in a future study.

\section{Damping of the oscillation and Reynolds number}\label{S:damp}

So far we concentrated on the frequency of the loop oscillation to investigate the magnetic field strength in the loop. We now turn to the damping time $\tau$ of the oscillation as defined through \equ{E:osci} and discuss consequences for the dissipation and thus for the (magnetic) Reynolds number.

Damping is commonly found in transverse loop oscillations. The damping time
is particularly important, because it reflects the rate at which the wave
energy is either converted into another wave mode or dissipated. Using a scaling relation derived
from numerical models, \citet{Nakariakov+al:1999} found that the magnetic
Reynolds number deduced from the observed damping time is of the order of $10^6$. This is several orders
of magnitude smaller than the classical value that would be of the order of $10^{14}$ and consequently would indicate an anomalously high dissipation. Later, \citet{Ofman+Aschwanden:2002} quantified this further through an analysis of the scaling relation between damping time and loop parameters. They estimated
an anomalously large viscosity of $10^{9.2\pm3.5}$\,m$^2$\,s$^{-1}$, which also gives 
a Reynolds number that is  much smaller than the classical
value. These results imply that the wave energy could be efficiently converted into heat
because of the anomalously high diffusivity.
Still, one has to keep in mind, that the damping is not necessarily due to dissipation of the wave energy (even though ultimately the wave will be dissipated). \citet{Goossens+al:2002} argued that the
observed fast damping of transverse loop oscillations can be explained by
the damping of quasi-mode kink oscillations. Thus, they suggested that there might be no need for the anomalously
large viscosity.

In our model the magnetic resistivity
$\eta$ is 5$\times$10$^{9}$\,m$^2$\,s$^{-1}$ (and the kinematic viscosity $\nu$ is of the same order).
Essentially this value is determined by the grid spacing in the numerical models \citep{Bingert+Peter:2011,Peter:2015} and ensures that currents that build up due to the driving of the magnetic field by photospheric motions will be dissipated at scales of the grid spacing. These values of resistivity and viscosity
are in the middle of the range of values found by \citet{Ofman+Aschwanden:2002}, i.e.\ $10^{9.2\pm3.5}$\,m$^2$\,s$^{-1}$. From this we conclude that the dissipation coefficient
 in our model (10$^{9.7}$\,m$^2$\,s$^{-1}$) is consistent
with the  high values found in  observations.

If we take the loop half width of our model loop (${\approx}1$\,Mm)
as a length scale, the dissipation time for the wave in our model would be $\tau_{\rm{diss}} \approx (1\,{\rm{Mm}})^2 / 5{\times}10^9\,{\rm{m}}^2\,{\rm{s}}^{-1} =   200\,{\rm{s}}$, which is
close to the damping time of 125\,s for the oscillation in our model. This suggests that  in
our model the damping is indeed due to (resistive and viscous) dissipation.
A further more detailed analysis of the energy budget will be needed
to investigate how much of the energy released by the trigger is first converted into the oscillation and then dissipated by viscosity and resistivity.
In particular further investigations will have to study the role of the spatial resolution and thus of the Reynolds number on the resulting oscillations.

\section{Conclusions}\label{S:sum}

We presented a loop oscillation found in a realistic coronal model driven by magnetic flux emergence through the photosphere (i.e. the bottom boundary). The treatment of the energy balance in the model allows us to synthesise imaging and spectroscopic observations from the model. There we found a  clear transverse loop oscillation that we identified as the fundamental fast kink mode. The period ($P{=}52.5$\,s) and damping time ($\tau{=}125$\,s) of the oscillation are consistent with observations. In particular, our oscillation also follows the observed relation between period and damping time deduced by \citet{Verwichte+al:2013b}. At least in our model the damping of the oscillation is due to resistive and viscous dissipation. To what extent this conclusion can be also drawn for the real Sun will have to be revealed by new simulations at higher spatial resolution.

We applied methods of coronal seismology  to the coronal emission synthesised from our model corona in the same way as it would be done for real observations. Here we concentrated on the oscillation in the Doppler shift of a coronal emission line. We chose a (vertical) line-of-sight for the analysis of the model data  that mimics the observation of a coronal loop near disk center.

Based on the fundamental kink mode oscillation we deduced an average magnetic field strength of $B_{\rm kink}{=}79$\,G in the loop.  In contrast to solar \emph{observations}, in the model we  know the magnetic field in the oscillating loop. This way we can understand what the field strength derived from coronal seismology actually means. In the coronal part of the model loop the magnetic field strength varies strongly by about a factor of five and drops to some 50\,G near the apex, i.e. significantly below the value derived from coronal seismology, $B_{\rm{kink}}$. The average of the field along the loops as suggested by \cite{Aschwanden+Schrijver:2011} gives a relatively good match, ${\langle}B{\rangle}{=}92$\,G. This average of the field corresponds to the constant field in a flux tube with a constant cross section that has the same wave travel time. Still, the difference between $B_{\rm{kink}}$ and ${\langle}B{\rangle}$ is considerable. Because many of the assumptions of the simple derivation of the wave period are violated in the model (and most certainly on the real Sun), e.g. the loop width, substructure, or dynamics, this is not too surprising.  

We conclude that the magnetic field strength deduced by coronal seismology can be a good representative of that in the upper part of coronal loop. Realistic models like the one presented here might further guide the interpretation of coronal oscillation results, in particular to relax some of the assumptions to get a better match of coronal properties derived from coronal seismology.  

\begin{acknowledgements}
{The authors thank Ding Yuan, Tom van Doorsselaere, and Leon Ofman for helpful suggestions. F.C. thanks Rony Keppens for supporting his visit to Katholieke Universiteit Leuven. The authors also thank Sven Bingert for the help on setting up the simulation, and Mark Cheung for offering the data that are used as the bottom boundary of the simulation. This work was supported by the International Max-Planck Research School (IMPRS) for Solar System Science at the University of G\"ottingen. We acknowledge PRACE for awarding us the access to SuperMUC based in Germany at the Leibniz Supercomputing Centre (LRZ).}
\end{acknowledgements}

\bibliographystyle{aa}
\bibliography{reference}

\end{document}